# Transverse spin and spin-orbit coupling in silicon waveguides

Alba Espinosa-Soria and Alejandro Martínez*

Nanophotonics Technology Center, Universitat Politècnica de València, Valencia, Spain

* email: amartinez@ntc.upv.es

**Abstract**: **Evanescent and tightly confined propagating waves exhibit a remarkable transverse spin density since the longitudinal component of the electric field is not negligible. In this Letter, we obtain via numerical simulations the electric field components of the fundamental guided modes of two waveguides typically used in silicon photonics: the strip and the slot waveguide. We obtain the relation between transverse and longitudinal field components, the transverse spin densities and other important parameters, such as the longitudinal component of the so-called Belinfante's spin momentum density (BSMD). By asymmetrically placing a circularly-polarized point-like dipole source in regions showing local circular polarization, the guided mode is excited unidirectionally via spin-orbit coupling. In contrast to metal plates supporting surface plasmons, the multimode behavior of silicon waveguides results in different spin-orbit coupling properties for each guided mode. Our results may find application in silicon photonic devices, integrated quantum optics and polarization manipulation at the nanoscale.**

It has been recently shown that, in contrast to propagating waves in free space exhibiting longitudinal spin, evanescent waves support transverse spin orthogonal to the wave vector [1-3]. This property of evanescent waves can be generally considered as a manifestation of the quantum spin Hall effect of light [4]. Such transverse spin is the ultimately responsible for the spin-controlled unidirectional excitation (SCUE) of guided waves [5-14], a phenomenon that has its origin in the spin-orbit coupling taking place when a circularly-polarized subwavelength dipole is placed in the evanescent-wave region of a guided mode. The spin-momentum locking inherent to evanescent waves [4,15] enables to completely switch the propagation direction of the excited guided mode by merely changing the spin of the exciting dipole [3,8-9]. By reciprocity, the spin of the wave scattered by a point-like defect placed in the evanescent-field region can be switched by simply reversing the direction of the guided wave [11,12]. This phenomenon, which takes place at any wavelength regime (for instance, at microwave frequencies [6,16]), may have especial relevance in nano-optics, since it ultimately enables mapping the light spin into a propagating pathway. Indeed, such spin-orbit coupling may induce lateral optical forces in order to manipulate a great amount of nanoparticles over large areas, as recently proposed [17].

Remarkably, longitudinal components of the electric field also take place in propagating fields when tightly confining light via strong focusing [18] or total internal reflection in dielectric structures such as microresonators [19] or waveguides [20,21]. The latter case is especially important, since high-index contrast waveguides made of semiconductor materials allow for an excellent field confinement leading to extreme miniaturization of photonic circuitry whilst keeping low propagation losses. In particular, silicon photonics, which relies on the confinement and guiding of light through silicon waveguides, is expected to become the mainstream technology for photonic integrated circuits. Although there have been some preliminary demonstrations of SCUE of guided modes in silicon waveguides [8,10], a detailed analysis of the electric field components of the guided modes and its relation spin-orbit related phenomena is, to our knowledge, still lacking.

Here, we numerically analyze the electric field components of the guided modes of two common silicon waveguides, the strip and the slot waveguide, at telecom wavelengths. We obtain cross-sectional maps of the relationship between transverse and longitudinal field components, the transverse spin densities and the longitudinal component of the so-called Belinfante's spin momentum density (BSMD) [18]. We show that, unlike surface plasmons propagating on infinite metal surfaces, in which the local polarization is always elliptical, silicon waveguide modes show points in which the projection of the electric field vector on a plane containing the propagation direction exhibit purely circular polarization. Moreover, such points can be found outside (in the evanescent region) as well as inside the waveguide core. By placing a circularly polarized point-like dipole source at such points, we show numerically SCUE of the guided mode via spin-orbit locking, being the propagating direction switched just by changing the spin of the exciting dipole. Remarkably, we show that each guided mode displays a different transverse spin both inside and outside the waveguide core. This results in completely different excitation conditions to achieve unidirectional guiding for each guided mode.

In our study, we consider first the silicon strip waveguide with rectangular cross-section, which is typically used in silicon photonics for guiding and processing light at telecom wavelengths. We assume that the silicon core is surrounded by air, although similar results could be obtained if other low-index materials such as silica or a polymer were considered as cladding. The waveguide axis is along the $z$-direction. Notice that, unlike in photonic crystal waveguides where the polarization distribution also changes along the waveguide axis [22], the strip waveguide exhibits continuous translational symmetry along this axis. Therefore, the polarization properties will not change with the $z$ coordinate so we restrict our study to the transverse $z=0$ plane. The propagating guided modes are characterized by three electric field components with a certain amplitude profile, $E_x(x,y)$, $E_y(x,y)$ and $E_z(x,y)$, and a real wave vector $k_z = 2\pi n_{\text{eff}}/\lambda$, where $n_{\text{eff}}$ is the effective index of the mode and $\lambda$ is the free-space wavelength. We consider the fundamental even mode, usually called TE-like mode, characterized by a strong electric field component along the $x$ direction. Indeed, this mode is typically excited in experiments by using an external light beam polarized in such direction. In waveguides with large-size core and low-index contrast, such as standard single-mode optical fibers, the longitudinal component of the electric field is negligible. However, as mentioned above, high-index contrast waveguides with strong confinement leads to considerable values of $E_z$. Unlike waveguides with cylindrical symmetry [9], there is not an exact analytical solution for the guided modes in the strip waveguide. Therefore, we have calculated numerically the fundamental even mode distribution by using the finite-element method (Synopsys FemSIM). In the calculations, we have considered a wavelength $\lambda = 1550$ nm, a grid size of 5 nm and a total simulation domain of 3x3 $\mu m^2$.

Figure 1(a) shows a cross-sectional map of the amplitude of the $z$-component of the electric field multiplied by the imaginary unit ($iE_z$) for propagation of the guided mode along $+z$. Throughout this work, we show the electric field components normalized with respect to the value of $E_x$ at the origin of coordinates. It can be seen that the $E_z$ field is asymmetric with respect to the $x=0$ plane, being its magnitude comparable to $E_x$ but with a phase shift of $\pi/2$ between them. To better appreciate the relation between the longitudinal and transverse components of the electric field, in Figs. 1(b) and (c) we plot the maps of $iE_z/E_x$ and $E_y/(iE_z)$ respectively. Notice that for each case we have chosen the order of the field components when performing the ratio in order to get a better visual representation. In contrast to surface plasmons on metallic surfaces or guided modes in cylindrical waveguides, in which the ratio between the longitudinal and transverse field components only depends to the distance to the interface, here we have a very inhomogeneous distribution, extremely different depending on the transverse field component we consider ($E_x$ or $E_y$). Remarkably, we find regions, both inside and outside the waveguide core, in which the longitudinal component is much higher (in absolute value) than the transverse ones. This behavior, which could be expected for the $E_y/(iE_z)$ case, is quite surprising for the $iE_z/E_x$ case, since $E_x$ is the fundamental component of the guided mode. It is also noticeable that maximum values of $|iE_z/E_x|$ are obtained mainly inside the waveguide core, this is, within the guided-wave region. Such high values take place in the core regions close to the lateral sidewalls as a result of the reduction of the $E_x$ component because of the strong index discontinuity along $x$.

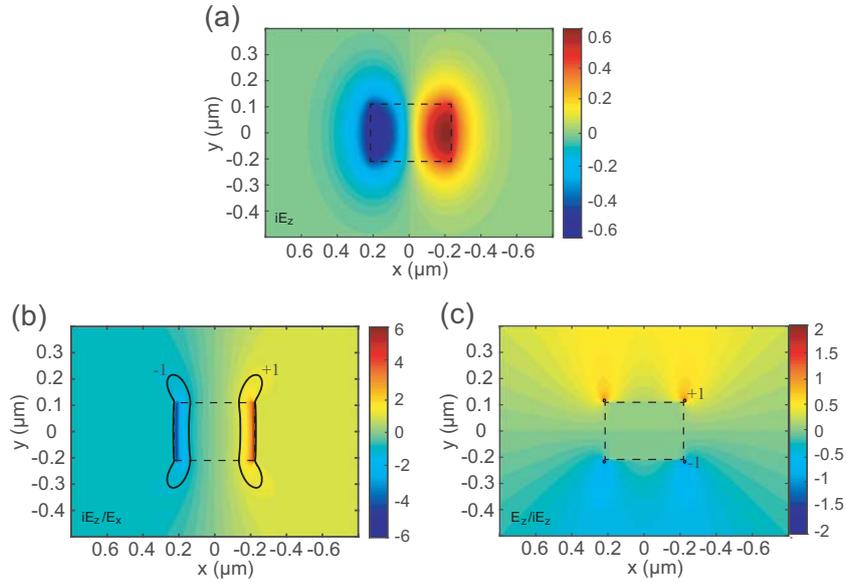

*Figure 1. Numerically calculated cross-sectional maps of (a) $iE_z$; (b) $iE_z/E_x$; and (c) $E_y/iE_z$ for the fundamental even mode of a silicon strip waveguide (with 450 nm x 220 nm cross-section) at $\lambda$ = 1550 nm. Contours with $iE_z/E_x$ or $E_y/iE_z$ = 1 and - 1, corresponding to points exhibiting local circular polarization of the field projected on the corresponding planes, are also highlighted. The dashed rectangle highlights the interface between the silicon core and the air cladding.*

From the electric field components, it is straightforward to obtain the electric contribution of the transverse spin densities of the guided mode as $s_E^x \propto Im(E_y^* E_z)$ and $s_E^y \propto Im(E_z^* E_x)$ [3,18]. The cross-sectional maps for both $x$ and $y$ spin densities (in normalized units) of the silicon strip waveguide are depicted in Figs. 2(a) and (b) respectively. In addition, we can also obtain the electric contribution to the longitudinal component of the so-called Belinfante's spin momentum density (BSMD) [23] as $p_{S,E}^z \propto (\partial_x s_E^y - \partial_y s_E^x)$ [18], which is plotted in Fig. 2(c). Notice the strong resemblance between $s_E^x$ (almost negligible), $s_E^y$ and $p_{S,E}^z$ along the $x$ axis for an $x$-polarized tightly focused beam (Fig. 3 in [18]) and our fundamental even in a high-index dielectric waveguide, as a result of the non-negligible longitudinal component of the electric field. However, the strong discontinuities of the field components perpendicular to the boundaries in the guided mode result in some differences, mainly the positive and negative $s_E^x$ spots near the waveguide corners and the existence of maximum values of $|p_{S,E}^z|$ at the waveguide boundaries for the guided mode.

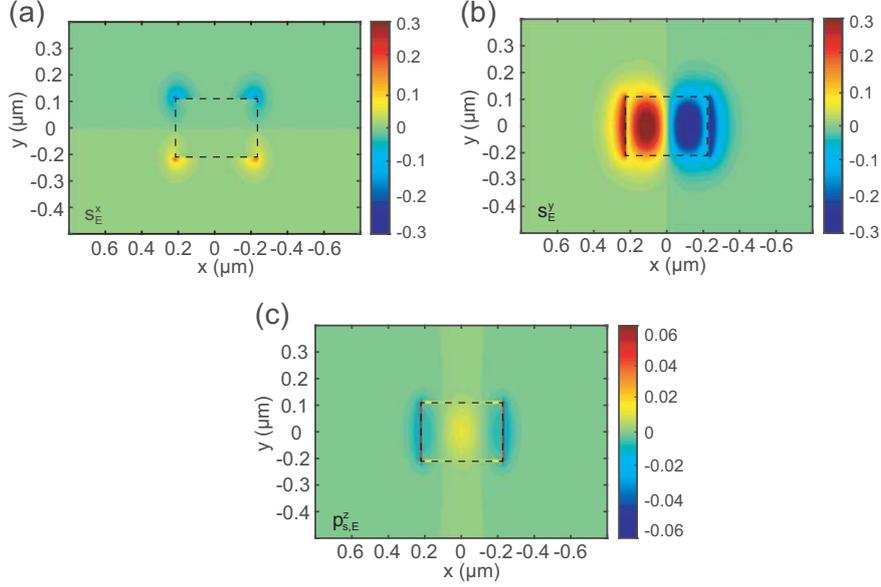

Figure 2. Numerically calculated cross-sectional maps of (a) $s_E^x$, (b) $s_E^y$, and (c) $p_{S,E}^z$ for the fundamental mode of the silicon strip waveguide at $\lambda = 1550$ nm.

In Figs. 1(b) and (c) we have highlighted contours in which $iE_z/E_x$ or $E_y/iE_z$ equals ±1 since such regions exhibit local circular polarization on the *xz* and *yz* planes (such points are commonly referred to as C-points [22])) associated to the guided even mode propagating along +*z*. This is a clear example of spin-momentum locking, which ultimately results in the SCUE of guided modes: when placing a point dipole emitting circularly polarized waves - or a point scatterer externally illuminated by a circularly polarized wave – at a point in such regions, the even guided mode will be excited towards one direction of the waveguide or the opposite depending on the sign of the spin of the excitation wave [10,13]. This idea is illustrated in Fig. 3(a) and checked via numerical simulations using a full 3D solver (CST Microwave Studio). The simulation has been performed using a time domain solver and hexahedral mesh with 10 cells per wavelength, except near the dipole where a refinement has been done (reaching 83 thousand cells in total). Open boundary conditions are chosen for all external faces. In Fig. 3(b) we show a snapshot of the propagating $E_x$ field on the *y*=0 plane when a point-like dipole with a dipolar moment $\vec{p} \propto \hat{y} + i\hat{z}$ is introduced at (*x,y*)=(1.39, 1.3)μm, where $iE_z/E_x = 1$. Notice that the even mode is clearly excited only towards one of the two possible directions. Such direction could be changed just by switching the spin of the exciting source. A similar result is obtained by placing a circularly polarized dipole ($\vec{p} \propto \hat{y} -$ $i\hat{z}$) at a spatial point in which $E_y/iE_z = 1$, as shown in Fig. 3(c). Now, the guide mode propagates towards negative *z* values, as the spin of the source is -1. This case is especially remarkable since the waveguide mode is unidirectionally excited by an optical source not containing any component along *x*, being $E_x$ the main polarization component of the guided mode under consideration. Simulations also show that the power contrast ratio between forward and backward directions is 17dB in the first case (b) and 25dB in the second case (c), which is a clear signature of unidirectional propagation. Nevertheless, even though very high values of the contrast ratio are achieved, the contrast ratio should be – ideally - infinite. We can explain the observation of a non-infinite contrast ratio by considering that diffraction and reflection of the launched waves in the waveguide boundaries will make the polarization at the dipole position not completely circular [9], resulting in a reduction of the unidirectionality.

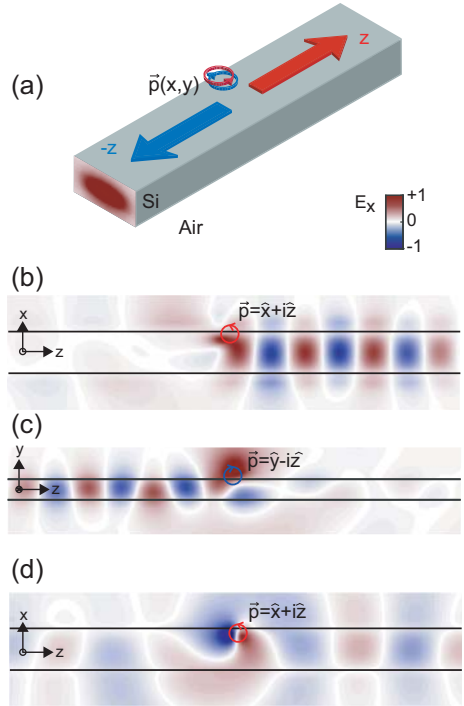

*Figure 3. (a) Scheme of the SCDE of the fundamental even mode in a silicon strip waveguide by using a circularly polarized point-like dipole with moment $\vec{p}$; Snapshots of the $E_x$ component (fundamental TE-like mode) for (b) $\vec{p} \propto \hat{x} + i\hat{z}$ placed at (x, y)=(1.39, 1.3)µm (y=0 cut) ; and (c) $\vec{p} \propto \hat{y} - i\hat{z}$ placed at (x, y)=(0.22, 1.21)µm (x=0 cut). (d) Snapshot of $E_y$ (TM-like mode) for (b) $\vec{p} \propto \hat{x} + i\hat{z}$ placed at (x, y)=(1.39, 1.3)µm (y=0 cut)*

Results shown in Figs. 1 and 2 describe the transverse spin behavior of the fundamental TE-like mode of a silicon strip waveguide, which is the mode typically employed to build silicon photonics functionalities and devices. Similar results – not shown here for the sake of simplicity- could be obtained for higher-order modes, if the waveguide core area is large enough to support them. These modes will display different profiles for the three electric field components. For instance, let us consider the second guided mode, usually termed TM-like mode and characterized by a main electric field component along the *y*-axis. The field components of this modes will resemble those for the TE-like mode (Figs. 1 and 2) but rotated 90º around the z-axis. Therefore, $E_x$ ($E_y$) for the TM-like mode will look similar to $E_y$ ($E_x$) for the TE-like mode, resulting in completely different conditions to get SCUE of the guide mode. This is checked in Figs. 3(d), which is similar to Fig. 3(b) in what refers to excitation conditions but depict the electric field component along *y* (and therefore, show the propagation of the TM-like mode). It can be seen that in these case the TM-like mode is not unidirectionally excited because the local polarization of the evanescent tails at the point where the dipole is placed is not completely circular but elliptical. This proves that spin-orbit coupling in multimode waveguides depends on the mode we are exciting, which has to be taken into account when using this effect to build nano-optical devices aimed at polarization manipulation.

Similar results can also be obtained for another kind of waveguide commonly employed in silicon photonics: the slot waveguide [24]. The slot waveguide comprises two parallel strip waveguides with a small gap between them. Remarkably, if the gap is sufficiently small, the fundamental even mode exhibits a strong $E_x$ confinement in the slot region, making this waveguide well suited for sensing [25] and all-optical processing [26,27]. Therefore, it makes sense to find out if such waveguide also present useful properties in what refers to the transverse spin of its fundamental mode, especially in the active slot region. Figure 4 shows the simulation results obtained for a slot waveguide. Since the slot waveguide can be considered as two coupled strip waveguides, many of the observations on polarization properties previously done for the strip case can also be applied here, mainly the existence of very high values of the transverse spin densities. The SCUE of the slot mode is demonstrated in Fig. 5 by using both *xz* and *yz* point-like circular dipoles. In both cases, the power ratio between forward and backward directions higher than 10 dB. Again, this is a signature of unidirectional propagation, but the value is not infinite because of the existence of diffracted and reflected waves that modify the polarization at the position of the excitation source. Remarkably, a guided mode with odd symmetry with respect to the *x*=0 plane (see the $E_x$ field pattern in the inset) propagating along negative *z* values is observed. This mode also arises from the splitting of the fundamental even mode of the isolated strip waveguide as a consequence of the strong coupling between parallel waveguides. As for the case of the strip waveguide, we see here that the condition for SCUE of the fundamental even mode only hold for that mode, and a higher-order mode is also excited without unidirectional behavior.

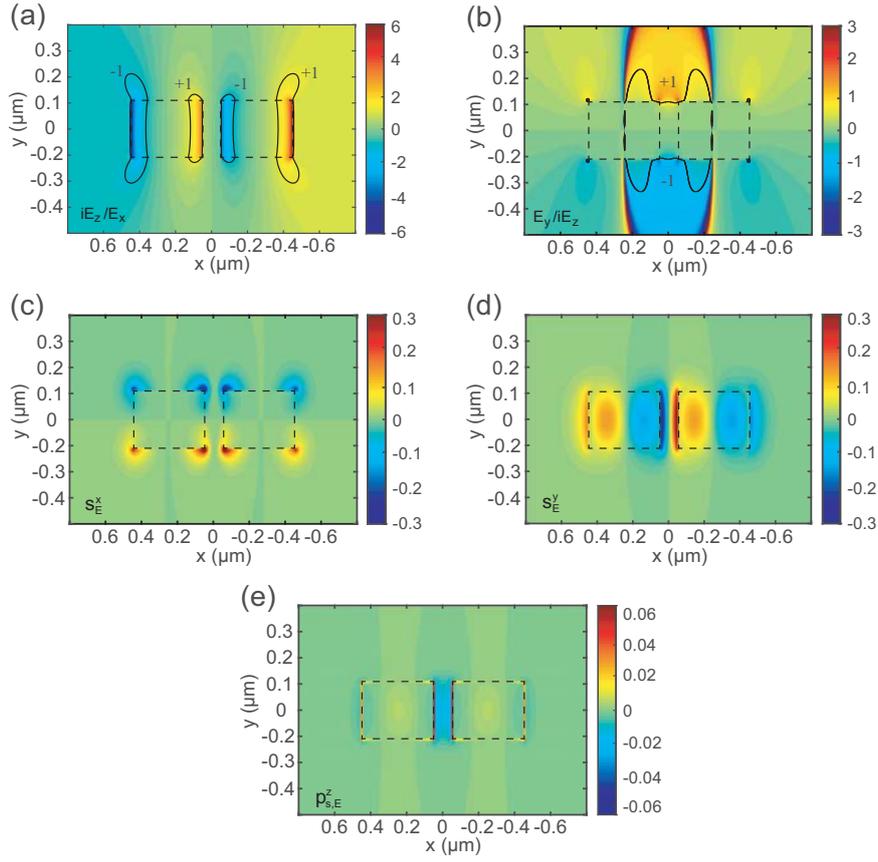

*Figure 4. Numerically calculated cross-sectional maps of (a) $iE_z/E_x$, (b) $E_y/iE_z$, (c) $s_E^x$, (d) $s_E^y$, and (e) $p_{s,E}^z$ for the fundamental mode of the silicon slot waveguide (comprising two strip waveguides with 300 nm x 220 nm cross-section spaced by a 100 nm gap) at $\lambda$ = 1550 nm.*

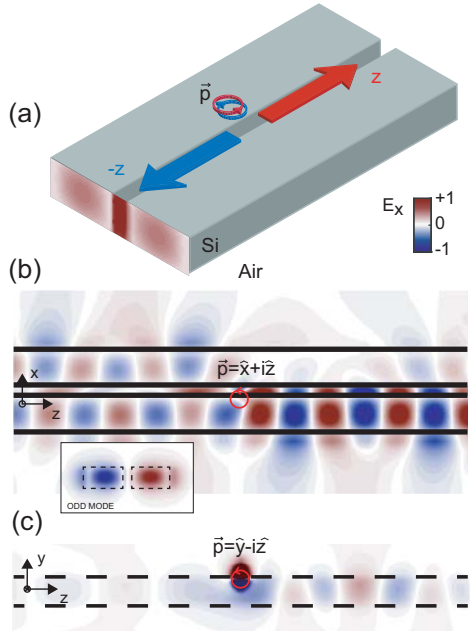

*Figure 5. (a) Scheme of the SCDE of the fundamental even mode in a silicon slot waveguide by using a circularly polarized point-like dipole with moment $\vec{p}$; Snapshots of the $E_x$ component for*

*(b) $\vec{p} \propto \hat{x} + i\hat{z}$ placed at (x, y)=(-0.59, 1.15 µm) (y=0 cut) and (c) $\vec{p} \propto \hat{y} + i\hat{z}$ placed at (x, y)=(0.01, 1.01) µm (x=0 cut).*

Summarizing, we have numerically analyzed the transverse spin as well as the orbit-coupling features in silicon strip and slot waveguides, providing cross-sectional maps that lock the position and polarization of a point-like source with the propagation direction of the excited guided mode. We have shown that the transverse spin can be significant not only in the evanescent region but also inside the waveguide core. Unlike more symmetric waveguides such as metal plates or cylindrical fibers, silicon waveguides exhibit local points exhibiting circular polarization in transverse planes both in the evanescent and guided regions. Importantly, the spin-momentum locking producing the SCUE of guided waves is different for each guided mode. Therefore, the unidirectional excitation for a certain excitation dipole will only take place for the mode satisfying the local polarization conditions imposed by the electric field components of the mode. This should be carefully taken into account in real applications, for instance, by engineering the waveguide to be single mode at the working wavelength or by externally filtering the undesired modes. Our results can be relevant for studying spin-orbit coupling effects (such as the SCUE of guided modes) in a photonic integrated platform, including not only silicon but also active III-V materials, with applications in multiple disciplines ranging from quantum processing to optical processing. In addition, we envisage that the existence of relevant transverse spin must be properly considered when coupling plasmonic metallic scatterers or nanoantennas to silicon waveguides [28,29,30].

**Funding.** Spanish Ministry of Economy and Competiveness (MINECO) (TEC2014-51902-C2-1-R); Valencian Conselleria d'Educació, Cultura i Esport (PROMETEOII/2014/034).

**Acknowledgement.** We thank F.-J. Rodríguez-Fortuño for helpful discussions.